\documentstyle[sprocl,epsfig]{article}

\begin{document}
\title{Atmospheric  neutrinos: phenomenological summary and outlook}
\author{Paolo Lipari}
\address{INFN and 
Dipartimento di Fisica, Universit\`a di Roma I }

\maketitle

\abstracts{
The predictions  of  the atmospheric $\nu$ event rates 
are affected by significant  uncertainties,
however the  evidence for the `disappearance' of
$\nu_\mu$'s and $\overline{\nu}_\mu$'s 
obtained  by SK  (and other underground  detectors)
is robust and cannot be  accounted in the framework of the 
minimum  standard  model
without assuming  very large   {\em ad hoc} experimental systematic  effects.
The existence  of `new physics'  beyond the  standard model
is  therefore   close to be established; $\nu$ 
oscillations   provide a  very   good  fit to all data.
The theoretical   uncertainties 
do have  an important  role  in the  detailed interpretation
of the data,   and  in the estimate  of  oscillation parameters.}

\section{Evidence for the disappearance of muon  neutrinos}
The data of  Super--Kamiokande  (SK) \cite{SK,SK1} and  other detectors
have   given  strong  evidence   that  $\nu_\mu$'s and $\overline{\nu}_\mu$'s
`disappear'.    
This   evidence  comes  from the  observation of  three
experimental effects:
(i)  the detection  of an up--down asymmetry
for the $\mu$--like  events,
(ii) the detection of  a small $\mu/e$ ratio,
(iii) the detection of  a  distortion   of the zenith angle
distribution and  a suppression of the $\nu$ induced upward 
going muon  flux.
The three  effects  are listed  in order of   `robustness'
with  respect to systematic  uncertainties.
The statistical significance  of the effects  in SK,
especially   for (i) and (ii),  is  very strong.
In the following we will discuss how theoretical 
uncertainties in the predictions
cannot `reabsorb'  the  observed   effects, 
but do play a role in the  {\em interpretation}  of the data.

\section{Systematic uncertainties in the predictions}
In the prediction of  the  event rates for 
an atmospheric $\nu$ experiment
one   needs to: (i) consider an  initial  flux of  cosmic ray particles,
(ii) model  the hadronic
showers  produced  by these particles in the Earth atmosphere,
(iii)  describe the $\nu$ cross sections,
(iv)  describe the  detector  response
to  $\nu$ interactions (and possible background  sources).
Here we  will consider   only the first  three `theoretical' elements of  
the calculation, and will  argue  that there are significant uncertainties,
that  influence the  absolute normalization, the 
shape of the   energy spectrum, the angular  distribution,
and the $\mu/e$  ratio  of the  events.
The  primary cosmic  ray  (c.r.) flux  has  been  a major  source
of  uncertainty  (see \cite{Gaisser} for a detailed  discussion)
because of   the discrepant\footnote{It is  highly unlikely that the
obseved differences are   the result  of  time  variations.}
results obtained  by two  groups (Webber 79  and LEAP 87) 
differing by  $\sim 50\%$ (see  fig.~1).
 Recently, new  measurements  of the  c.r. proton  flux
have  given results consistent  with the lower 
normalization.  
If the  lower normalization is  accepted  as  correct,
the uncertainty  in the primary  c.r. flux can be 
reduced, however the  description of 
the primary   c.r. flux  (see fig.~1) used   in 
two  calculations  of 
the atmopheric  $\nu$ fluxes: Honda et al.  (HKKM)  \cite{HKKM} and 
Bartol \cite{Bartol} used  in predictions for SK
and  other  detectors  are  then  too  high (by $\sim 30\%$ and 
$\sim 10$\%).   
\begin{figure} [bt]
\centerline{\psfig{figure=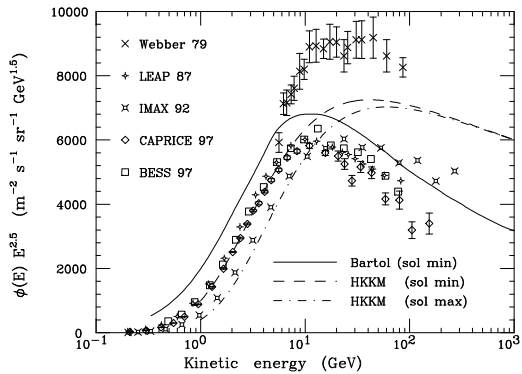,height=5cm}}
\caption {Measurements of the c.r. proton flux  (points)
(see \protect\cite{Gaisser} for the references), and the representations
used as  input in the  Bartol \protect\cite{Bartol}  and
HKKM  \protect\cite{HKKM} calculations (lines).}
\end{figure}

A second  source important  source of uncertainty is our lack of
knowledge  on the properties of particle  production 
in   $p$--nucleus  and nucleus--nucleus  interactions.
The calculations  of Bartol \cite{Bartol} and HKKM \cite{HKKM} use  different
descriptions for the multiplicity and
energy spectrum  of  the pions produced 
in $p$--Nitrogen (Oxygen)  interactions.  
For the same  primary c.r.  flux,  this  would  result
in a $\sim 20\%$  higher  $\nu$ event  rate 
for the Bartol  calculation.
Some  controversy  exists  about  what   description 
of hadronic  interactions is  in  better agreement with the existing data.
An  experimental    program,
studying   in detail  the structure
of  particle  production  in the  relevant   energy range 
(a  broad region centered  at $E_0 \sim 20$~GeV),
would  result  in an   improvement in the  predictions.

The similar normalization of the  two \cite{Bartol,HKKM} calculations
is   the  result  of a cancellation  between  
a  higher (lower) flux for the c.r.  flux, 
and  a lower (higher)  $\nu$ yield  per primary particle
for the  HKKM (Bartol)  calculation.
This  cancellation, to a large extent, is not
casual, but is the  consequence  of  fitting  
the (same) data on $\mu^\pm$ fluxes  at  ground level.
This   underlies  the importance  of   these measurements.
It  is very desirable to  repeat  them with  greater   accuracy.
High altitude  meaurements
with balloons \cite{Gaisser}  also offer  a great potential.

A third  source of `theoretical' uncertainty, of  comparable  importance to
the other  two, is  related
to  the description of $\sigma_\nu$.
At  high   $E_\nu$, when most of the   phase space
for  $\nu$ interactions  is in  the deep--inelastic   region,
$\sigma_\nu$ is   reliably calculable in  terms  of  well  determined
parton distribution  functions (PDF's).  However   for $E_\nu \sim 1$~GeV
the  description of  $\sigma_\nu$ is  
theoretically   more difficult.  Quasi--elastic scattering  is
the  most  important  mode, but also   events  with  the production 
of one or  more  pions  (where  the additional  particles are
undetected  or  are  reabsorbed in the  target  nucleus)  are important
contributions  to the  signal.  The  production 
of  $\Delta$'s  and   other  resonances  is
important, nuclear   effects    have to be  included.
A  relatively   small  modification in the description of
a   fraction of $\sigma_\nu$ in the SK  montecarlo:
the choice of  a  new set  of  PDF's
(GRV94LO  replacing  the CCFR parametrization) has     resulted in an  increase
of the predicted  number of  partially contained  events by 
 approximately 7\%
(compare the  MC predictions  in \cite{SK} and \cite{SK1}).
It appears  very  difficult  to  calculate  accurately 
$\sigma_\nu$ from first principles in the 
relevant  energy region.  The existing  data     do  not 
determine  the   absolute  value of  the cross section 
and the energy spectrum of the final state lepton better that 
$\sim 15\%$.   Additional data 
could help in improving  the situation.  The 
K2K $\nu$ beam with a spectrum not too different
from the  atmospheric one offers    interesting  possibilities.

\section {Robust  properties  of the predictions and  observed  effects}
Two  properties  of  the $\nu$ fluxes
are   to a large extent  independent from the details  of the 
calculation   and  provide  `self calibration'    methods:
(i) the fluxes are approximately up/down  symmetric:
$\phi_{\nu_\alpha}  (E_\nu, \theta)   \simeq  
\phi_{\nu_\alpha}  (E_\nu, \pi - \theta)$,
(ii) the  $\nu_\mu$  and $\nu_e$ fluxes
are strictly related  to each other  because they are
produced  in the   chain decay of the same
charged  mesons (as in $\pi^+ \to \nu_\mu \, \mu^+$  followed
by $\mu^+ \to \overline{\nu}_\mu \nu_e e^+$). Writing 
$\phi_{\nu_\mu} (E,\theta) = r (E_\nu, \theta) \times \phi_{\nu_e} (E,\theta)$  
the  factor $r (E_\nu, \theta)$  varies  slowly with
energy and  angle  and is quite insensitive   to   the details  of 
 the calculation.
These two properties  are  at  the  basis  of the   robustness of the
evidence for oscillations.
The up--down symmetry  follows  as  a simple 
and {\em  purely geometrical}  consequence   from two  assumptions:
the  primary c.r.  flux  is isotropic,
the Earth is  spherically symmetric.
The  c.r. flux  at  1 A.U. of  distance form   the sun 
is isotropic  to  a   precision   better than $10^{-3}$,
as can be measured observing in a fixed  direction  and  
looking  for time   variations   while the Earth  rotates.
The isotropy is spoiled by the geomagnetic   field   that
bends  the particle  trajectories  and  forbids   the lowest
rigidity  ones from reaching the Earth's  surface introducing directional
(east--west)  and  location    (latitude) effects.
\begin{figure} [bt]
\begin {minipage} [h] {15cm}
\psfig{figure=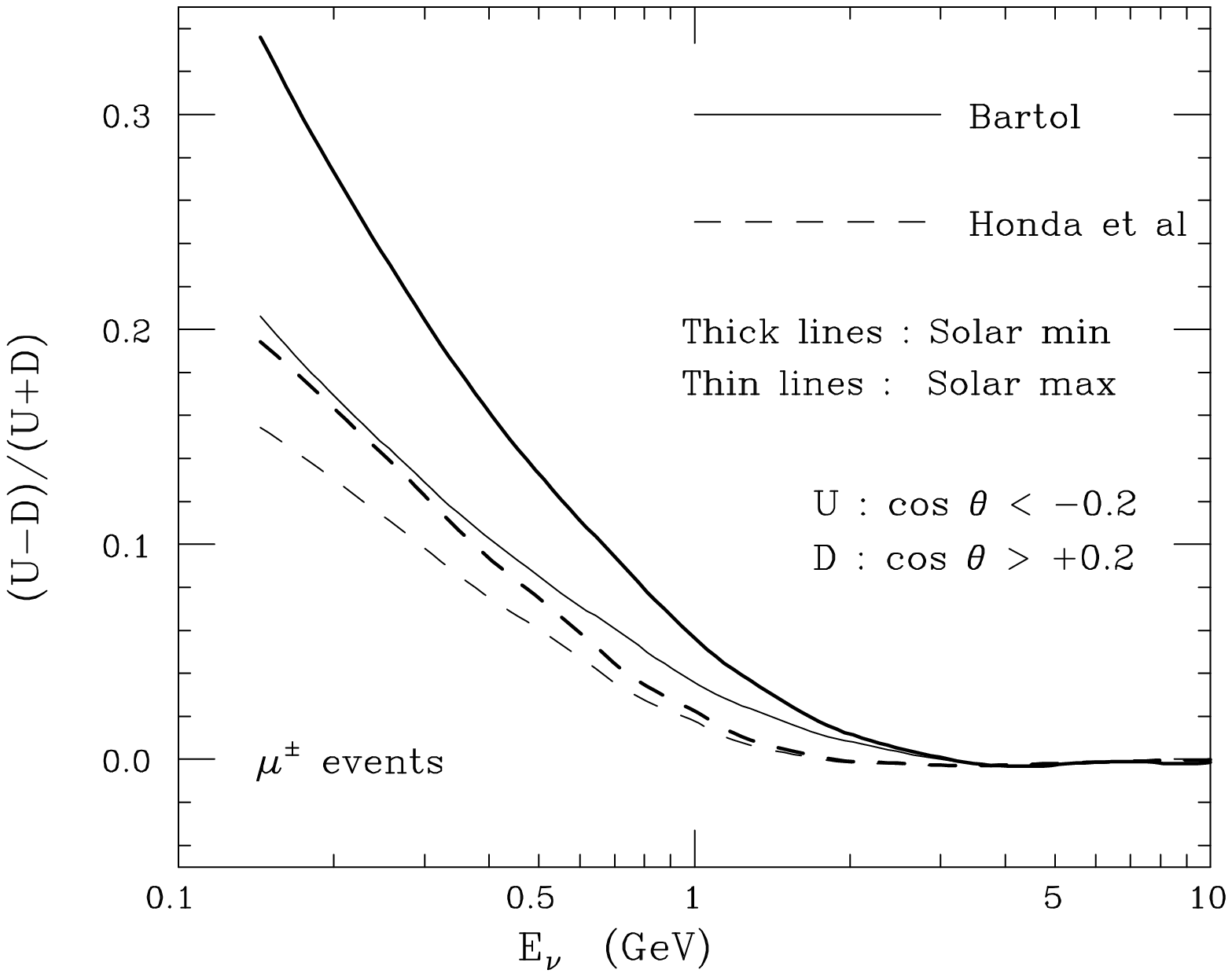,height=4.4cm}
\psfig{figure=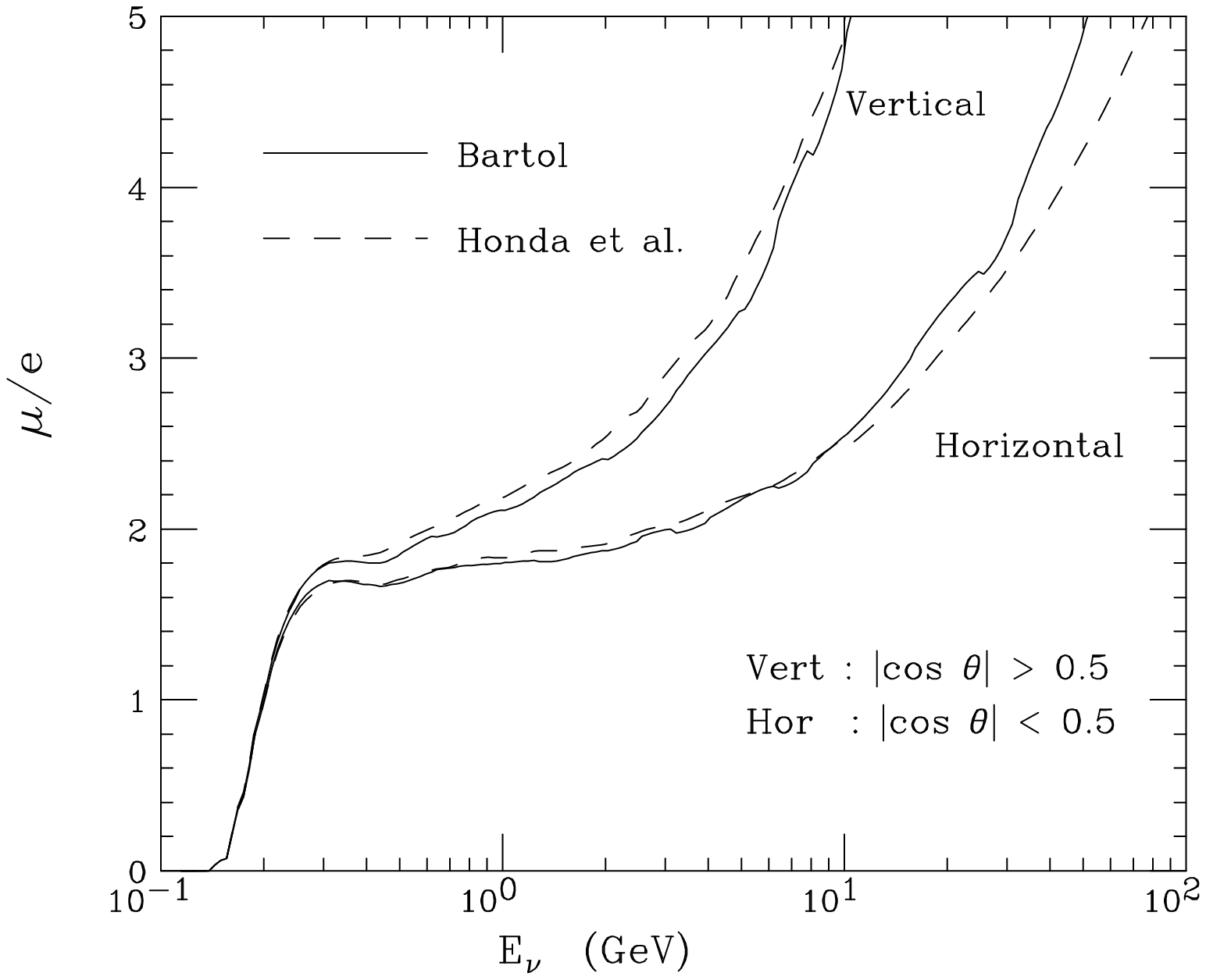,height=4.4cm}
\end{minipage}
\caption {The  left panel  shows
the up/down asymmetry  (no--oscillation hypothesis) 
of cc--interacting $\nu_\mu$'s
and $\overline{\nu}_\mu$'s  in  the Bartol \protect\cite{Bartol}
and  HKKM \protect \cite{HKKM} calculations for  the Kamioka site.
The right panel  shows the $\mu/e$ ratio for  the two models.}
\end{figure}
These effects vanish  at   large momentum (see fig.~2).
The  measurement of an (oscillation independent) 
east--west effect for  atmospheric $\nu$'s
in agreement with predictions, by SK \cite{SK-east-west}
is  an important test, that  validates the calculations.
Note  that at Kamioka  (near the magnetic  equator)
geomagnetic  effect have the opposite effect to  $\nu$--oscillations,
and  produce  an up--going $\nu$ flux
{\em larger}   than the down--going one 
(both magnetic  poles  are  {below}
the detector). 
At the Soudan mine (near the magnetic  pole)
the opposite  is  true. 
Note also that the predicted asymmetry  at low  energy 
 (see fig.~2)  has some model dependence  with the Bartol calculation
predicting  a higher  no--oscillation  asymmetry.
This  is  important for the 
detection of  a zenith  angle  modulation in the Soudan   detector
and  is also   relevant for the interpretation of the  SK  sub--GeV  events.

The right  panel  of  fig.~2  shows  how   different
calculations of  the atmospheric  $\nu$ flux  predict  very similar
$\mu/e$  ratios. 
This  however  refers   to a {\em  fixed}  value of $E_\nu$.
In fig.~3 
\begin{figure} [hbt]
\centerline{\psfig{figure=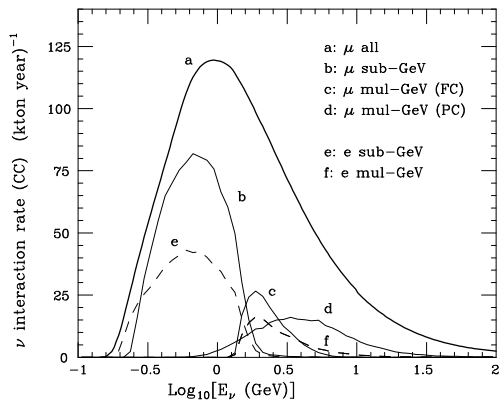,height=5.0cm}}
\caption {
Approximate energy  distributions
of  interacting neutrinos  for different
classes of events in SK. Note the difference
in  the shape of the response for $\mu$--like  and  $e$--like   events  }
\end{figure}
we show an estimate of  the energy  distributions  of the neutrinos
that  produce  the SK events,   note  how the distributions  for 
$\mu$ and $e$ like  events   differ.
For the multi--GeV samples, 
a  harder  $\nu$ spectrum,  or a  faster  raise with energy of 
$\sigma_\nu$, result in  a    larger  (smaller) 
increase  in the predicted  rate of $\mu$--like ($e$--like) events,
therefore in a smaller double  ratio  $R$  (because of a 
larger  denominator)  and   finally to  a larger  $\Delta m^2$ 
(for  the same  mixing) in  the  $\nu_\mu \leftrightarrow  \nu_\tau$  
interpretation  to explain the larger  suppression.
At this  conference,  SK has  presented
a  new  estimate
for the (90\% C.L.) allowed  region in the 
($\sin^2  2 \theta$, $\Delta m^2$)  plane for  the 
$\nu_\mu \leftrightarrow  \nu_\tau$    hypothesis, 
considering  a  larger  exposure  and 
a  slightly modified  MC calculation.
The  new \cite{SK1}  allowed region  is  smaller that the
previously  published  one \cite{SK} not  including  the    
interval $|\Delta m^2| \simeq 0.5$--$1.0 \times 10^{-3}$~eV$^2$,
a   result    very encouraging for the LBL  programs.
The  use  of a  new set  of PDF's in  the description of $\sigma_\nu$
has the qualitative  effect  to  enhance      the contribution  of
high  energy events,   and   for the argument  outlined  above
is  an  important  contribution to the exclusion of the low 
$\Delta m^2$  interval.

\section{Outlook}
The  detection of  oscillations  in   atmospheric $\nu$ 
experiments  is a result of  great importance.
The detailed  study of this  phenomenon  
and   the precise measurement  of the  parameters 
(masses and  mixing)  involved is  a great  opportunity
and a difficult challenge.
SK has a  remarkable  potential  to 
obtain  more  convincing  evidence and more  precise measurements.
New  data  on primary c.r.  fluxes, hadron--nucleus  interactions, 
$\nu$--nucleus  interactions  and $\mu^\pm$ fluxes  could  help 
the interpretation of present  and  future data.
Long  baseline  $\nu$ beams,   have  also  the
potential  to  confirm the   results
and  study the  phenomenon.
This could   happen very soon with the K2K project,
the existence of two  (similar  to each other) LBL projects  in the
US  and in Europe is  seen by  some  as  a beneficial  case 
of scientific competition, and  by  others  as  a   dangerous waste of
resources.

\noindent {\bf Acknowledgments:} Special thanks to prof. Y.~Suzuki
for kind explanations.

\end{document}